\pdfoutput=1
\documentclass[
reprint,
superscriptaddress,
nofootinbib,
amsmath,amssymb,
aps,
prd,
]{revtex4-2}

\usepackage[table,svgnames,dvipsnames]{xcolor}
\definecolor{Blue}{HTML}{1c1e94}
\usepackage[normalem]{ulem}
\usepackage{aas_macros}
\usepackage{graphicx}
\usepackage{dcolumn}
\usepackage{bm}
\usepackage[mathlines]{lineno}
\usepackage{orcidlink}
\usepackage{cases}
\usepackage{booktabs}
\usepackage{comment}
\usepackage{multirow}
\usepackage{makecell}
\usepackage{siunitx}
\usepackage{booktabs}
\graphicspath{{figs/}}
\usepackage{fontawesome5}
\usepackage{hyperref}

\def\be{\begin{equation}}
\def\ee{\end{equation}}
\newcommand{\vs}{\nonumber\\}
\def\ba#1\ea{\begin{align}#1\end{align}}
\newcommand\numberthis{\addtocounter{equation}{1}\tag{\theequation}}

\newcommand{\code}[1]{{\texttt{#1}}}
\renewcommand{\emph}[1]{\textit{#1}}

\usepackage[nameinlink,noabbrev]{cleveref}
\crefname{equation}{Eq.}{Eqs.}
\crefname{section}{Section}{Sections}
\crefname{figure}{Fig.}{Figs.}
\crefname{table}{Table}{Tables}
\crefname{appendix}{Appendix}{Appendices}
\Crefname{figure}{Figure}{Figures}
\Crefname{equation}{Equation}{Equations}
\Crefname{section}{Section}{Sections}
\Crefname{table}{Table}{Tables}
\newcommand{\refeq}[1]{Eq.~(\ref{eq:#1})}
\newcommand{\refeqs}[2]{Eqs.~(\ref{eq:#1})--(\ref{eq:#2})}
\newcommand{\reffig}[1]{Fig.~\ref{fig:#1}}

\newcommand{\refapp}[1]{Sec.~\ref{app:#1} (SM)}

\newcommand{\refsm}[1]{Sec.~\ref{app:#1} (SM)}
\newcommand{\refsmb}[1]{Sec.~\ref{app:#1}}

\newcommand{\<}{\left\langle}
\renewcommand{\>}{\right\rangle}

\newcommand{\LL}{\mathcal{L}}
\renewcommand{\L}{\Lambda}
\renewcommand{\P}{\mathcal{P}}

\newcommand{\U}{\mathcal{U}}

\newcommand{\N}{\mathcal{N}}

\newcommand{\fli}{\mathrm{FLI}}

\newcommand{\obs}{\mathrm{obs.}}

\renewcommand{\max}{\mathrm{max}}

\newcommand{\kmax}{k_\max}

\newcommand{\eps}{\epsilon}
\newcommand{\sigmaEps}{\sigma_\eps}
\DeclareMathOperator{\tr}{tr}

\newcommand{\Del}{\mathcal{D}}

\newcommand{\shat}{\hat{s}}

\renewcommand{\d}{\delta}
\newcommand{\dg}{\delta_g}

\newcommand{\Plin}{P_{\mathrm{L}}}

\newcommand{\dlin}{\delta^{(1)}}

\newcommand{\Ngrid}{N_{\mathrm{grid}}}
\newcommand{\NgridEul}{N^{\mathrm{Eul}}_{\mathrm{grid}}}

\def\bOset{\{b_O\}}
\def\dgdet{\d_{g,\rm det}}
\def\dgrec{\d_{g,\rm rec}}
\def\dlinL{\dlin_{\Lambda}}

\newcommand{\Mpch}{\,h^{-1}\mathrm{Mpc}}

\newcommand{\iMpch}{\,h\,\mathrm{Mpc}^{-1}}

\newcommand{\Msunh}{\,h^{-1} M_\odot}

\newcommand{\vx}{\bm{x}}
\newcommand{\vk}{\bm{k}}
\newcommand{\vq}{\bm{q}}

\begin{document}

\title{Forward vs Backward: Improving BAO Constraints with Field-Level Inference}

\author{Ivana Babi\'{c}} \email{I.Babic@physik.uni-muenchen.de}
\affiliation{Max–Planck–Institut für Astrophysik, Karl–Schwarzschild–Straße 1, 85748 Garching, Germany}
\author{Fabian Schmidt\orcidlink{0000-0002-6807-7464}} \email{fabians@mpa-garching.mpg.de}
\affiliation{Max–Planck–Institut für Astrophysik, Karl–Schwarzschild–Straße 1, 85748 Garching, Germany}
\author{Beatriz Tucci\orcidlink{0000-0003-2971-2071}} \email{tucci@mpa-garching.mpg.de}
\affiliation{Max–Planck–Institut für Astrophysik, Karl–Schwarzschild–Straße 1, 85748 Garching, Germany}

\date{\today}

\begin{abstract}
We present results of field-level inference of the baryon acoustic oscillation (BAO) scale $r_s$ on rest-frame dark matter halo catalogs.
Our field-level constraint on $r_s$ is obtained by explicitly sampling the initial conditions along with the bias and noise parameters via the \code{LEFTfield} EFT-based forward model.
Comparing with a standard reconstruction pipeline applied to the same data and over the same scales, the field-level constraint on the BAO scale improves
by a factor of $\sim 1.2-1.4$ over standard BAO reconstruction.
We point to a surprisingly simple source of the additional information.
\end{abstract}

\maketitle

\textit{Introduction.}---The baryon acoustic oscillation (BAO) feature is currently one of the prime science targets of galaxy redshift surveys.
This feature is so important because it allows us to measure the distance-redshift relation $d_A(z)$ as well as the Hubble rate $H(z)$ as a function of redshift, thus probing the evolution of dark energy, as well as the mass of neutrinos. The unique large-scale oscillatory pattern of the BAO feature (of order a $100\,\rm Mpc$ in comoving units) allows for these robust measurements by avoiding significant degeneracies with other cosmological parameters, as well as the bias parameters describing the relation between galaxies and matter on large scales. Conversely, the broad-band shape of galaxy clustering as a function of scale probes the evolution of the growth of structure in the Universe, as well as the above-mentioned bias parameters. Importantly, the BAO scale measurement is strictly relative to $r_s$, the sound horizon at the baryon drag epoch, which is well constrained assuming standard early-Universe evolution, but could be modified in the presence of early dark energy or additional relativistic degrees of freedom before recombination. So, keeping the late-time expansion history constrained, one can turn the BAO scale measurement around into a constraint on $r_s$ and thus early-Universe physics. The potential of BAO  measurements is highlighted by the recent DESI DR2 measurements, demonstrating distance constraints relative to $r_s$ at the level of 0.5--1\% \cite{DESI-DR2}.

While the BAO feature, thanks to its shape and large-scale nature, is largely preserved in the late-time galaxy distribution, it is both damped and slightly shifted by the nonlinear evolution of large-scale structure. In particular, the large-scale displacements of matter and galaxies under gravity essentially smooth out the feature.
The damping effect leads to a loss in the statistical precision of the BAO measurement from galaxy clustering, because it reduces the contrast of the feature. Moreover, the small systematic shift induced in the inferred scale is no longer negligible given the substantial volume covered by current galaxy sureys.

In order to deal with these issues, the current state of the art is to perform a so-called \emph{BAO reconstruction}, where the large-scale displacements responsible for the BAO damping and shift are estimated from the galaxy distribution itself \cite{Rec_Eisenstein_2007, Eisenstein_2007_rec_2, Padmanabhan_2009_rec}. Then, the observed galaxy positions are moved by the opposite of the estimated displacement, thus attempting to reverse the damping effect on the BAO feature. Finally, the BAO scale is estimated from the power spectrum of the galaxy density field measured in terms of the displaced positions.
The detailed algorithm is described in \refsmb{reco} in the supplementary material (SM).
While elegant and easy to implement, the standard BAO reconstruction approach suffers from several drawbacks. First, the displacement estimate is necessarily based on assumptions of a fiducial cosmology and a value for the linear galaxy bias parameter. Second, standard BAO reconstruction does not remove the effects of nonlinear galaxy formation (nonlinear bias), which, as we will see, also play a role in the precision of the BAO scale measurement. Finally, the displacement is estimated from the galaxy density field only at linear order (Zel'dovich approximation); improvements on this approximation have been suggested in the literature \cite{Baojiu_rec,Noh_2009_reconstruction,  Tassev_2012_reconstruction, Burden_2015_reconstruction, Schmittfull_2015_reconstruction, Wang_2017_reconstruction, Schmittfull_2017_reconstruction, modi/feng/seljak:2018, McQuinn:2020, 2023MNRAS.520.6256S,2023MNRAS.523.6272C,parker/bayer/seljak:2025}, but so far have not been applied to large redshift surveys.

Here, we present an alternative approach to BAO inference that addresses all of these shortcomings: \emph{field-level Bayesian inference} (FLI). This approach jointly estimates the BAO scale, large-scale displacements, and bias parameters in a consistent Bayesian framework, by explicitly sampling the initial conditions, i.e. the linear matter density field $\dlin(\vk)$, where the BAO feature is imprinted. More precisely, the BAO scale enters the inference via the structure of the prior on the initial conditions, which is determined by the linear power spectrum $\Plin(k|r_s)$ which depends on $r_s$ [\refeqs{deltalin_shat}{fdef}]. The accuracy of the BAO inference is then limited only by the accuracy of the \emph{forward model} for the observed galaxy density field, $\d_g[\dlin]$ (the results presented here will be in fact on simulated dark matter halo catalogs). 
The latter is in fact likely to be the key challenge for field-level inference \cite{Nguyen:2020hxe,Villanueva-Domingo:2022rvn,Kostic:2022vok}: since we cannot simulate realistic galaxy catalogs in a cosmological volume, and certainly not at a computational cost that would allow for field-level inference, model misspecification is inherent in any forward model, and the robustness of the results against such misspecification must be validated. Unlike the standard reconstruction approach, field-level inference uses the entire available data set, and a field-level inference of the BAO scale that is robust to model misspecification is a nontrivial demand. 

The forward model employed here, \code{LEFTfield} \cite{Schmidt:2020ovm},
is based on a Lagrangian formulation of the Effective Field Theory (EFT) of LSS \cite{Baumann:2010tm,Carrasco:2012cv,Carroll:2013oxa}.
This model incorporates the dependence of the galaxy density on all local observables of a comoving observer, while enforcing 
Einstein's equivalence principle which ensures that matter and galaxies co-move on large scales \cite{Desjacques:2016bnm}.
Indeed, this fact allows for the inference of large-scale displacements, and thus is the basis also for the standard BAO reconstruction.
Unlike the model employed in standard BAO reconstruction, our forward model includes all contributions up to third order in perturbation theory, and is able to predict the halo density field accurately up to wavenumbers of $k\sim 0.2 \iMpch$ \cite{Schmidt:2020viy,Schmidt:2020tao,Babic:2022dws,stadler/etal:2024a}.

We should highlight previous studies which performed field-level inference of distances from large-scale structure. Ref.~\cite{Ramanah:2018eed} presented an inference of cosmology via Alcock-Paczy\'nski distortions on mock data generated from the forward model, demonstrating substantial improvement over the power spectrum without reconstruction.
Recently, Ref.~\cite{babic/etal:2024} demonstrated field-level BAO-scale inference using \code{LEFTfield} on mock catalogs generated by the EFT model at a higher scale than used in the inference, showing robustness to model mismatch.

Here, we demonstrate for the first time that the model can successfully perform  field-level inference of the BAO scale on fully nonlinear tracers: dark matter halos in N-body simulations. 
Given that our forward model is agnostic to the details of the tracer considered---and based only on the equivalence principle---we expect that our results generalize to actual galaxies as well (see \cite{2021JCAP...08..029B,Beyond2pt:preprint} for field-level results on simulated galaxies).

\medskip

\textit{Data.}---We consider two halo samples. The SNG sample consists of main halos in the $\log_{10} M_{200m}=12.5-14.0\Msunh$ mass range, identified with the \code{ROCKSTAR} halo finder \cite{Behroozi:2011ju} at redshift $z=0.50$ in an N-body, gravity-only simulation. This sample has a mean comoving number density of $\bar n= 1.3\cdot10^{-3}(\Mpch)^{-3}$. The simulation assumes a flat $\Lambda$CDM cosmology, encompasses a comoving volume $L^3=(2000\Mpch)^3$ and contains $N_{\mathrm{particle}}=1536^3$ particles of mass $M_{\mathrm{particle}} = 1.8\times10^{11}\Msunh$ \cite{Schmidt:2020tao}. 
The second, Uchuu halo sample, consists of main halos in the $\log_{10} M_{200m}=12.0-13.5\Msunh$ mass range, identified with \code{ROCKSTAR} at redshift $z=1.03$ ($\bar n= 3.6\cdot10^{-3}(\Mpch)^{-3}$) in the Uchuu simulation \cite{Ishiyama:2020vao}, which assumes a different flat $\Lambda$CDM cosmology. This simulation spans the same volume $L^3=(2000\Mpch)^3$ while offering a much higher mass resolution, $N_{\mathrm{particle}}=12800^3$ particles of mass $M_{\mathrm{particle}} = 3.27\times10^{8}\Msunh$. Notice that the initial conditions for the Uchuu simulations were not available to the authors. 
Throughout, the model is evaluated at the fixed redshift of the respective sample, and we drop time and redshift arguments for clarity in the following.

\medskip

\textit{Field-level EFT of biased tracers.}---The EFTofLSS provides a perturbative framework within which the galaxy or halo density field $\dg$ can be systematically expanded, order by order in perturbations, as
\be
\dg[\dlin](\vk)=\sum_O b_O O[\dlin](\vk) + \eps(\vk).
\label{eq:deltag_def}
\ee
At each given order, there is a finite number of galaxy bias operators $O[\dlin]$, each associated with a coefficient $b_O$.
This includes $b_\d \d[\dlin](\vk)$, where $\d[\dlin]$ is the \emph{forward-evolved} matter density, which is the well-known linear relation that is employed in the standard BAO reconstruction.
Notably, \refeq{deltag_def} includes several additional nonlinear bias terms, namely the complete set of 6 bias operators at second and third order; see \refsm{bias} for the detailed description. Note that the EFT counterterms for matter itself are a subset of the bias terms and hence already included in \refeq{deltag_def}.
The bias operators are jointly computed with the nonlinear evolution, for which we adopt second-order Lagrangian perturbation theory (2LPT).

The second term on the r.h.s. of \refeq{deltag_def} is the stochastic contribution encapsulating the random nature of small-scale fluctuations and the discrete nature of galaxies. These are described by the noise field $\eps$, which to leading order is Gaussian with RMS
\be
\sigma_\eps(k) = \sigma_{\eps,0}\left[1 + \sigma_{\eps,k^2}k^2\right],
\label{eq:sigmaEps_def}
\ee
where the subleading contribution $\propto k^2$ captures the finite extent of galaxy-forming regions.

\medskip
\textit{Field-level forward model.}---We forward model the bias fields $O$ in \refeq{deltag_def} starting from Gaussian initial conditions parametrized via a unit Gaussian random field $\shat\sim\N(0,1)$, discretized on a grid of size $\Ngrid$. The linear density field is related to $\shat$ by
\be
\dlin(\vk,z|r_s)=f(k,r_s)\,\left[\frac{\Ngrid^3}{L^3}\Plin^{\rm fid}(k,z)\right]^{1/2}\,\shat(\vk),
\label{eq:deltalin_shat}
\ee
where $L$ is the simulation box side length, $\Ngrid$ is the size of the
initial conditions grid, and $\Plin^{\rm fid}(k,z)$ is the linear power spectrum
in the fiducial (simulation) cosmology. We implement a change in the BAO scale
$r_s$ via the function $f(k,r_s)$ defined by
\ba
f(k,r_s) &= \frac{T^2_{\rm BAO}(k| r_s)}{T^2_{\rm BAO}(k|r_{s,\rm fid})}\,, \vs
T^2_{\rm BAO}(k|r_s) &= 1+ A\sin (k\,r_s+\phi)\exp(-k/k_{\rm D})\,,
\label{eq:fdef}
\ea
where $r_{s,\rm fid}$ is the comoving sound horizon in the fiducial cosmology (see \refsm{bao_param} for how we determine the value of $r_{s,\rm fid}$ and of the transfer function parameters $A$, $\phi$ and $k_{\rm D}$). 
Thus, unlike what is done in BAO scale inferences on actual data, where
the angular diameter distance to a given redshift is varied, 
we instead vary the BAO scale in the linear density field.
We adopt this approach, since it enables a change in the BAO scale while retaining the periodic boundary
conditions within the simulation volume, which is a significant technical
simplification. We stress that the exact same approach for varying the BAO
scale is used in both the field-level and standard reconstruction analyses; thus the
comparison is consistent, and we expect it to be an accurate indicator for what
would be obtained when varying the distance instead.

Our forward model employs the Lagrangian, EFT-based forward model of cosmological density fields, \code{LEFTfield} \cite{Schmidt:2020ovm}.
Following the EFT principle, \code{LEFTfield} evolves all cosmological (plus auxiliary) fields up to a finite EFT cutoff $\Lambda$ \cite{Carroll:2013oxa,Schmidt:2018bkr,Schmidt:2020viy}. Specifically, we implement a cubic sharp-$k$ filter on the initial conditions.
\code{LEFTfield} computes both $O=O(\shat)$ and $\partial O/\partial\shat$, the latter of which is essential for the Hamiltonian Monte Carlo technique employed. 
We refer to \refapp{fwd} and \cite{Schmidt:2020ovm,Schmidt:2020viy,Kostic:2022vok,stadler/etal:2024a} for details on the implementation and validation.
The forward model adopted here corresponds to the third-order Lagrangian bias forward model presented as alternative in the Supplementary Material of \cite{fbisbi}, where comparable posteriors on the parameter $\sigma_8$ were reported from both Eulerian and Lagrangian bias expansions. The Lagrangian bias expansion avoids an additional cut on the evolved density field, yielding improved numerical convergence properties \cite{stadler/etal:2024a}, and thus is preferred.

\medskip
\textit{Field-level BAO inference.}---In the field-level Bayesian inference pipeline, we evaluate and sample from an explicit field-level likelihood $\LL_\fli$. 
Following \cite{Cabass:2020nwf}, our analyses assume Gaussianity of galaxy stochasticity and analytically marginalize over $\eps$. 
This leads to a Gaussian log-likelihood for an observed galaxy field $\d^\obs_{g}$ of the following form \cite{Schmidt:2018bkr,Cabass:2019lqx}:
\ba
&\LL_\fli\left(\d^\obs_{g}  \Big|\shat, r_s, \{b_O\}, \{\sigma_{\epsilon,i}\}\right) =
-\frac{1}{2}\sum_{\vk \neq \bm{0}}^{|\vk| < \kmax} \label{eq:FLI_likelihood}\\
&\qquad\Big[\ln{2\pi\sigmaEps^2(k)}
+\frac{1}{\sigmaEps^2(k)}
\Big\lvert
\d^\obs_{g}(\vk) - \sum_O b_O O[\shat, r_s](\vk)
\Big\rvert^2
\Big]\,\,.\nonumber
\ea
The $\sum_{\vk \neq \bm{0}}^{|\vk| < \kmax}$ amounts to a spherical sharp-$k$ filter which only includes Fourier modes of the data up to $\kmax$. We also exclude the $\vk=\bm{0}$ mode, as it is degenerate with the effective mean density of the sample. 
For the analyses presented here, we mostly adopt $\kmax = \L$ (we will also discuss results with $\L > \kmax$), but note that the initial conditions filter $\L$ is cubic and thus includes modes up to $\sqrt{3} \L$. 
We expand $\sum_Ob_OO$ to third order in the galaxy bias operators $O$, and further analytically marginalize over all bias coefficients except $b_\d$ (this is possible thanks to the simple dependence of \refeq{FLI_likelihood} on the $b_O$ \cite{Elsner:2019rql}) assuming weakly informative Gaussian priors (see \refapp{Priors}). 

The final explicitly sampled FLI parameter space consists of
$\{ r_s, \{\shat\},b_\delta,\sigma_{\eps,0}, \sigma_{\eps,k^2}\}$,
where $\shat$ is a three-dimensional grid of size $\Ngrid^3$.
To explore this high-dimensional posterior, following \cite{Jasche:2018oym,Kostic:2022vok}, we employ two MCMC sampling methods: Hamiltonian Monte Carlo (HMC) \cite{Neal:HMC} for $\{\shat\}$---leveraging the differentiability of the \code{LEFTfield} forward models---and slice sampling \cite{Neal:slice} for
$\{r_s,b_\delta,\sigma_{\eps,0}, \sigma_{\eps,k^2}\}$.
The posterior validation, effective sample size, and correlation lengths are
presented in \refapp{validation}.

\medskip

\textit{BAO inference from the post-reconstruction power spectrum.}---For comparison and benchmarking, we aim to replicate a standard BAO reconstruction pipeline, taking care not to include modes beyond the cutoff scale $\kmax$ considered in the FLI analysis. For this purpose, we assign the tracer catalog to a density grid using the same grid resolution and assignment scheme used in the FLI analysis. We then filter the assigned tracer density with a Gaussian filter on the scale $R = \Lambda^{-1}$ before estimating the displacement. After re-assigning the tracer positions to their estimated Lagrangian positions, we evaluate the post-reconstruction power spectrum $P_{\text{p-rec}}(k)$ up to the same scale $\kmax=\L$ used in the FLI. We fit a model including a broad-band part parametrized with 5 free parameters, and obtain the best-fit and error bar on $r_s$ via Monte Carlo sampling using a Gaussian power spectrum likelihood. The details are described in \refsm{reco}.

We note that we fix the linear bias $b_1$ in the displacement estimation, and any uncertainty on $b_1$ is not propagated into the error on $r_s$. Further,
the power spectrum likelihood assumes a Gaussian covariance, and as such 
ignores any additional correlations induced by mode coupling (this approximation has been validated for DESI in Ref.~\cite{forero-sanchez/etal}) . In contrast, both of these sources of uncertainty are incorporated in the posterior on $r_s$ obtained in the FLI analysis. 
In \cite{babic/etal:2024} we have validated both analyses on mock data, and showed that standard reconstruction and FLI agree where they should, namely in the case of a linearly biased tracer of the 1LPT (Zel'dovich) density field with fixed linear bias.

\begin{figure}[t]
   \includegraphics[width=0.8\linewidth]{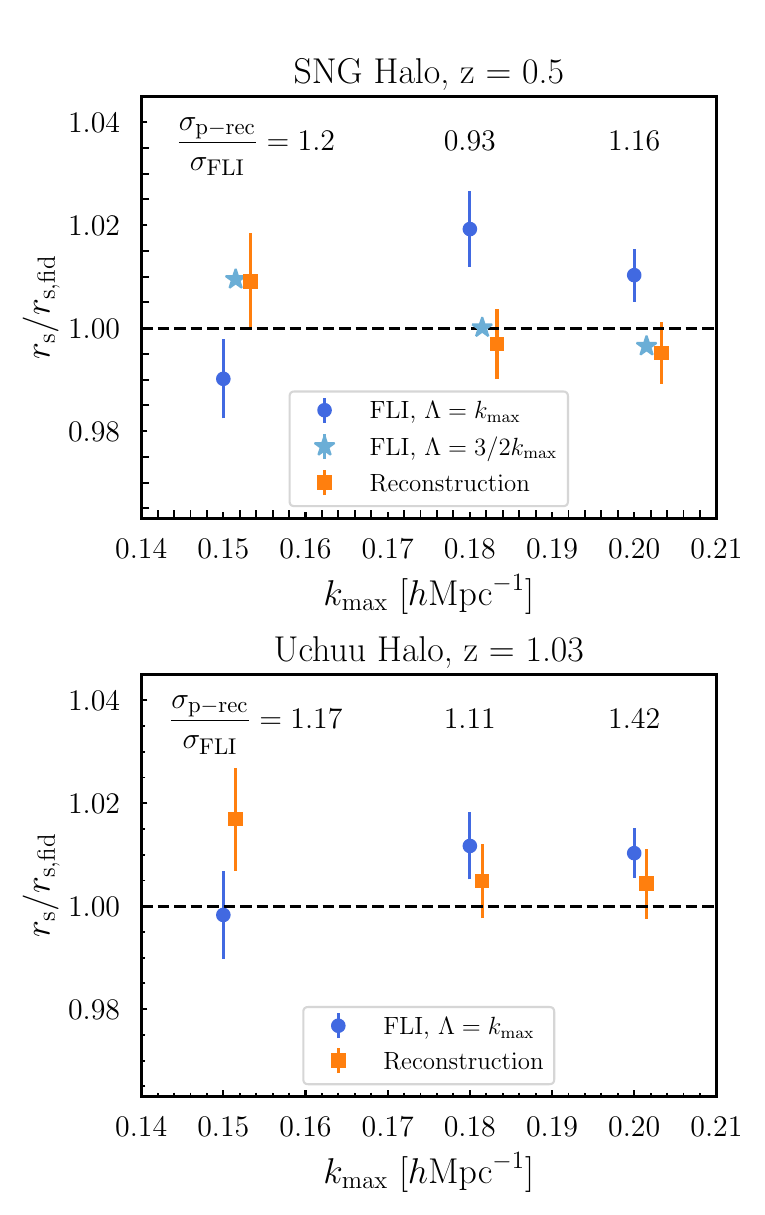}
   \caption{
Inferred BAO scale using field-level inference (FLI, blue circles) and post-reconstruction power spectrum (orange boxes) as a function of $\kmax$. The upper panel shows results for the SNG halo sample, while the lower shows the Uchuu halo sample. We also indicate the ratio in BAO scale uncertainties from both methods; FLI increases the BAO scale precision by 10--40\%, with the exception of the intermediate cutoff in the SNG case. The blue stars show FLI results using the same scales of the data, but with increased $\Lambda$ (see text for discussion).}
   \label{fig:beta_constraint_summary}
 \end{figure}

\medskip

\textit{Results.}---Our main results are shown in \reffig{beta_constraint_summary}, where we show, as a function of the maximum wavenumber included in the analysis, the inferred posterior mean and RMS for $r_s$ relative to the fiducial value $r_{s,\rm fid}$. Unbiased inference thus corresponds to a value consistent with 1.
While the standard reconstruction approach yields unbiased BAO constraints, FLI yields slightly biased results for the two higher cutoffs in the SNG sample. This bias can be traced back to our choice of $\Lambda=\kmax$. To ensure that the EFT bias terms can completely absorb the $\Lambda$-dependence of the model, one should ideally choose $\Lambda \gg \kmax$ \cite{Rubira:2023vzw,MAP_paper}, which however is numerically costly. We have subsequently performed inferences choosing $\Lambda=(3/2)\kmax$ for the SNG sample. The results (\reffig{beta_constraint_summary}) indicate that this choice removes the systematic shift in $r_s$ \footnote{Due to the limited sample size, we do not show the inferred error bar for these inferences, but we expect them to be very similar to the $\L=\kmax$ cases.}. FLI results for the higher-redshift Uchuu sample show a smaller systematic bias, which we expect will likewise be improved when increasing $\L/\kmax$.

\begin{figure}[t!]
            \includegraphics[width=0.9\linewidth]{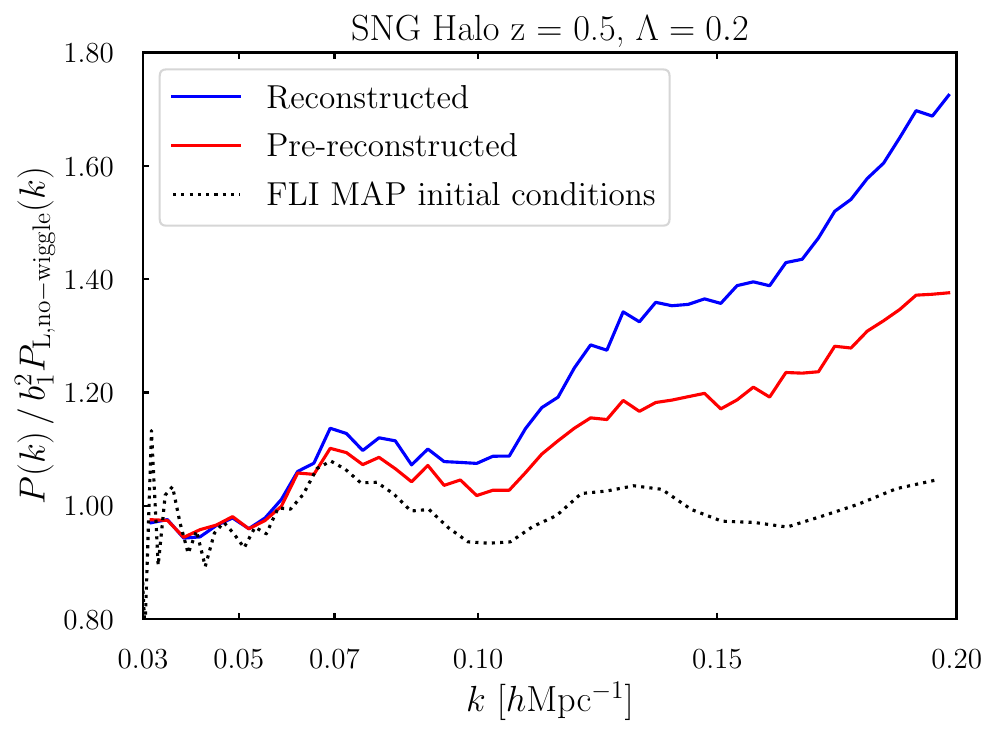}
            \caption{
            Ratio of the pre- and post-reconstruction halo power spectrum of the SNG sample to the smooth, ``no-wiggle'' part of the linear power spectrum multiplied by the best-fit $b_1^2$. The BAO feature is enhanced post-reconstruction, but there is a significant broad-band enhancement over the linear power spectrum. On the other hand, FLI reconstructs the linear density field and thus ``measures'' the BAO scale without a broad-band enhancement. This is illustrated by the power spectrum of the maximum-a-posteriori initial conditions sample from one of the SNG FLI chains (green dotted). 
}
            \label{fig:Pk-over-PLsm}
 \end{figure}

In \reffig{beta_constraint_summary} we also indicate the ratio of the 68\% CL error on $r_s$ between the post-reconstruction power spectrum analysis and the FLI result.
Despite taking fully into account the reconstruction uncertainties and marginalizing over all EFT bias terms, FLI generally improves over the BAO scale uncertainty 
in the standard reconstruction analysis, for both halo samples, by 10--40\%. The only exception is the intermediate cutoff for the SNG sample, where FLI is also most significantly biased. This data point should be revisited with the $\Lambda> \kmax$ inference in the future. Notice that the absolute error bars depend sensitively on $\kmax$, and current analyses typically use larger $\kmax$ values than presented here. However, we expect our conclusions to continue to hold, or become even stronger, at higher $\kmax$ (see \cite{fbisbi}), provided we remain within the range of validity of the EFT approach.

\medskip
\textit{Where does the additional information come from?}---The information gain demonstrated in \reffig{beta_constraint_summary} can have several causes. 
We investigate this analytically by considering the limit of infinitely informative (zero-noise) data following \cite{Cabass:2023nyo,MAP_paper}.
\refapp{Fisher} derives the Fisher information on the BAO scale in both FLI
and reconstruction approaches in this simplified setting. The result is
\begin{widetext}
\ba
F_{r_sr_s}^{\rm FLI} = -\<\frac{\partial^2}{\partial r_s^2} \ln \P_{\rm FLI}[\bOset, r_s | \d_g]\bigg|_{r_{s,\rm fid}}\> 
   =\:& 
\frac12 \sum_{\vk}^\L\frac1{[\Plin(k|r_{s,\rm fid})]^2} \left(\frac{\partial \Plin(k|r_{s,\rm fid})}{\partial r_{s,\rm fid}}\right)^2
\label{eq:FisherFLI}\\
F_{r_sr_s}^{\rm rec-P(k)} = 
-\<\frac{\partial^2}{\partial r_s^2} \ln \P_{\rm rec-P(k)}[r_s | \d_g]\bigg|_{r_{s,\rm fid}}\> 
   =\:&
\frac12 \sum_{\vk}^\L
\frac{1}{[ P_{\text{p-rec}}(k|r_{s,\rm fid})]^2}
\left(\frac{\partial P_{\text{p-rec}}(k|r_s)}{\partial r_s}\right)^2\,.
\label{eq:FisherPkrec}
\ea
\end{widetext}
Comparing \refeq{FisherPkrec} and \refeq{FisherFLI} we can identify the following factors that contribute to increased information on the BAO scale at the field level:
\begin{enumerate}
\item \textit{Improved reconstruction quality:} in FLI, the inferred initial density field $\dgdet^{-1}[\d_g, \bOset]$ is a representative sample from the Gaussian distribution characterized by the linear power spectrum $\Plin(k|r_s)$, provided that the forward model correctly describes the data on the range of scales considered. The inferred initial density field  thus carries the full BAO information encoded in $(\partial\Plin/\partial r_s)^2$. On the other hand, in the standard reconstruction-based analysis, this is replaced by $(\partial P_{\text{p-rec}}/\partial r_s)^2$, which could be suppressed relative to $\partial\Plin/\partial r_s$, for example because the 1LPT-based reconstruction does not restore the BAO wiggles as well as the 2LPT forward model employed in FLI.
To test this, we have also performed an inference with the 1LPT forward model for matter, keeping all aspects of the forward model including the bias expansion the same. This case yields very similar constraints on $r_s$ as the fiducial inference using the 2LPT forward model, which indicates that reconstruction quality is not the dominant effect on the scales considered here. 
\item \textit{Reduced variance:} the variance of each mode in the FLI is controlled by $\Plin(k|r_{s})$. This is illustrated by the maximum-a-posteriori (MAP) sample of the initial conditions in the FLI analysis shown in \reffig{Pk-over-PLsm}, which differs from the denominator $P_{\rm L,no-wiggle}$ only via the BAO oscillations. On the other hand, the variance in the standard reconstruction analysis is determined by $P_{\text{p-rec}}(k|r_{s})$. Since the standard reconstruction approach only reverts large-scale displacements, $P_{\text{p-rec}}$ is not close to $\Plin$, but has a substantially increased broad-band part due to contributions from nonlinear evolution and nonlinear bias (notice that the linear bias drops out of the zero-noise Fisher information in both cases). This broad-band contribution is clearly visible in \reffig{Pk-over-PLsm} (shown there for the SNG sample; the corresponding figure for the Uchuu sample is shown in \reffig{Pk-over-PLsm-Uchuu} (SM)). In fact, this increased variance, estimated from \reffig{Pk-over-PLsm}, can by itself explain at least a significant part of the observed information gain.
The non-Gaussian contributions to the covariance of $P_{\text{p-rec}}$ neglected in \refeq{FisherPkrec} generally work in the same direction.
\end{enumerate}
The simplified Fisher analysis above did not consider that the BAO feature in the post-reconstruction power spectrum is extracted via template fitting. The additional free parameters in the template can in principle further reduce the information over that in \refeq{FisherPkrec}.
\reffig{Corner_plot_rec} indicates that this should not be a dramatic degradation, but could be noticeable at the $\sim 10\%$ level.

One might wonder then if it is possible to improve the standard BAO reconstruction to also remove the broad-band contribution, i.e. to make $P_{\text{p-rec}}$ as close as possible to $\Plin$. However, this is not easy to do, since the broad-band shape arises from nonlinear mode coupling across a broad range of scales due to both nonlinear evolution and nonlinear bias. It would thus not only require a full inverse-model of galaxy clustering, but also information on the nonlinear bias parameters. FLI on the other hand jointly infers the higher-order bias parameters (essentially using higher-order $n$-point functions of the data \cite{MAP_paper}) with the BAO scale, so that the power spectrum of the inferred initial density field is indeed very close to $\Plin(k|r_s)$ (\reffig{Pk-over-PLsm}).
This conclusion is also supported by the test on linearly biased 1LPT mocks presented in \cite{babic/etal:2024}.

\medskip

\textit{Summary and discussion.}---Field-level inference of the BAO scale is the unique method that allows for a consistent inference of cosmological distances and parameters while reaping the benefits of BAO reconstruction. As such, it is worth pursuing even if it yields comparable error bars to standard reconstruction approaches, which keep cosmological parameters in the reconstruction step fixed. Here, we have presented the first field-level BAO inference on fully nonlinear tracers. We have found that, compared to a standard BAO reconstruction pipeline using the same scales of the data, an improvement on the BAO uncertainty by up to 40\% is achievable. The more accurate forward model employed in FLI enables several sources of information gain. The likely dominating factor for our results here is the incorporation of nonlinear bias (coupled with nonlinear evolution), which is responsible for a significant contribution to the post-reconstruction broad-band power spectrum and hence error bar.

In this context, it is worth discussing the assumption of a fixed broad-band linear power spectrum shape in the analysis here, which, it should be stressed, is assumed consistently in both FLI and reconstruction analyses. In an application to real data, one would exploit the fact that the FLI analysis implements a consistent Bayesian inference framework, and would thus jointly vary all cosmological parameters that describe the linear power spectrum as well as the distance-redshift relation in the analysis. Such an inference would consistently combine BAO, broad-band, and growth-of-structure information.

One important constraint of the EFT-based FLI analysis is that it is restricted to perturbative scales. On the other hand, standard BAO analysis often fit the BAO feature to smaller, nonlinear scales, which are no longer under perturbative control. Still, we stress that we have not yet fully exploited the range of scales accessible to the perturbative treatment in the results presented here, especially at redshifts $z\gtrsim 1$. Moreover, when pushing to smaller yet perturbative scales, the improved BAO sharpening due to the higher-order gravity forward model could lead to additional improvements. It would further be interesting to validate our analysis on simulated galaxy, rather than halo catalogs. Even more importantly, the inclusion of redshift-space distortions following \cite{stadler/etal:2024b} is not only essential for the application to real data, but could lead to further gains of field-level BAO constraints over standard reconstruction. We will pursue this in upcoming work.

\medskip

\begin{acknowledgments}
\textit{Acknowledgments.}---We thank
Andrija Kosti\'{c}, Minh Nguyen, Martin Reinecke, Ariel S\'{a}nchez, Julia Stadler, Uro\v{s} Seljak
for helpful discussions. 
FS acknowledges discussions and collaborations with \href{https://www.aquila-consortium.org/}{Aquila consortium} members on previous works that eventually led to these results.

Our analyses are performed on the \code{freya} and \code{orion} clusters, maintained by the \href{https://www.mpcdf.mpg.de/}{Max Planck Computing \& Data Facility}, as well as the \code{ada} cluster hosted and maintained by MPA.
We acknowledge the non-negligible \href{https://www.mpcdf.mpg.de/about/co2-footprint}{carbon footprint} of computational research and associated environmental impacts.
Our post-processing pipeline is empowered by \href{https://root.cern.ch}{\code{ROOT}}, \href{https://root.cern.ch/manual/python/}{\code{pyROOT}}, \href{https://getdist.readthedocs.io/en/latest/}{\code{GetDist}}, \href{https://numpy.org/}{\code{numpy}}, \href{https://emcee.readthedocs.io/en/stable/}{\code{emcee}}, and \href{https://matplotlib.org/}{\code{matplotlib}}.

\end{acknowledgments}

\bibliographystyle{apsrev4-2}
\bibliography{references}

\newpage
\pagebreak
\appendix
\onecolumngrid

\widetext
\begin{center}
\textbf{\Large Supplementary Material}
\end{center}

\section{Forward model}
\label{app:fwd}

The \code{LEFTfield} forward model is detailed in \cite{Schmidt:2020ovm} and \cite{stadler/etal:2024a}.
Here, we describe the specific procedure to construct the evolved matter density field $\d$ and bias fields $O$ out of a particular realization of the linear density field $\dlin$:
\begin{enumerate}
\item Apply a cubic sharp-$k$ filter with cutoff $\Lambda$ to the linear density field $\dlin \to \dlin_\Lambda$ in Fourier space. 
\item Construct the Lagrangian displacement field $\bm{s}(\vk)=\sum_n\bm{s}^{(n)}(\vk)$, where $n$ is the order of Lagrangian Perturbation Theory (LPT); in this work, we adopt $n=2$, i.e. 2LPT.
\item At the same time, construct the bias operators (see \refsmb{bias}).
\item Displace pseudo-particles from Lagrangian to Eulerian positions by moving them by the displacement field evaluated at their position. Each pseudo-particle carries a set of weights corresponding to each of the Lagrangian bias operators.
\item Assign the pseudo-particles to the Eulerian grid of size $\NgridEul$ via a non-uniform-to-uniform fast Fourier transform (NUFFT \cite{NUFFT}), using the set of weights to generate Eulerian bias operators $O(\vx)$. A unit weight is used to generate the evolved 2LPT matter density field $\d(\vx)$.
\end{enumerate}

The fields are effectively filtered with a spherical sharp-$k$ filter at the scale $\L$ in the evaluation of the likelihood in \refeq{FLI_likelihood}.
The detailed choices made in the forward model were investigated and validated in \cite{stadler/etal:2024a}.

\subsection{Galaxy bias expansion}
\label{app:bias}

The EFT galaxy bias expansion expands the local galaxy density field $\dg$ in a set of galaxy bias operators, i.e. fields, constructed recursively out of second derivatives of the gravitational potential. There are different options for choosing a complete basis of operators at a given order in perturbation theory, in particular whether one uses evolved (Eulerian) fields or initial (Lagrangian) fields. The Lagrangian approach offers several advantages: it allows a straightforward construction of higher-order bias operators using recursion relations \cite{Mirbabayi:2014zca,Schmidt:2020ovm}; by displacing these operators using the LPT solution, we automatically include all relevant higher-order displacement terms; and they allow for a direct incorporation of redshift-space distortions \cite{schmittfull+:rsd,stadler/etal:2024b}.

Here, we thus adopt the Lagrangian bias approach. The downside of this formulation is that the costliest step, the assignment of densities from a set of pseudo-particles, has to be repeated for each bias operator. Thanks to the short correlation length of the BAO scale, the increased cost of this forward model remains manageable (for reference, the previous study of field-level inference of the $\sigma_8$ parameter required more than a factor of 10 more samples for converged results \cite{fbisbi}).

The starting point for the Lagrangian bias expansion is the distortion tensor, defined as
\be
M^{(n)}_{ij}(\vq,\tau) \equiv \partial_{(i} s^{(n)}_{j)} (\vq,\tau),
\ee
where $\bm{s}^{(n)}$ is the $n$-th order contribution to the displacement in LPT introduced above.
Up to third order, the complete set of Lagrangian bias operators is given by \cite{Mirbabayi:2014zca,Desjacques:2016bnm}
\ba
&\text{1st-order} \, &&\tr\left[\bm{M}^{(1)}\right] \quad\Leftrightarrow\quad \d\vs
&\text{2nd-order} \, &&\tr\left[\bm{M}^{(1)}\bm{M}^{(1)}\right],\ 
\tr\left[\bm{M}^{(1)}\right]\tr\left[\bm{M}^{(1)}\right], \label{eq:biaslist} \\
&\text{3rd-order} \, &&\tr\left[\bm{M}^{(1)}\bm{M}^{(1)}\bm{M}^{(1)}\right],\ 
\tr\left[\bm{M}^{(1)}\bm{M}^{(1)}\right]\tr\left[\bm{M}^{(1)}\right],\ 
\tr\left[\bm{M}^{(1)}\right]\tr\left[\bm{M}^{(1)}\right]\tr\left[\bm{M}^{(1)}\right],\ 
\tr\left[\bm{M}^{(1)}\bm{M}^{(2)}\right]\,.
\nonumber
\ea
In practice, we replace the 1st-order Lagrangian bias operator with the Eulerian matter density $\d(\vx,\tau)$, such that its coefficient can be straightforwardly identified with the usual linear bias $b_\delta = b_1$.
Finally, we also include the leading higher-derivative bias operator,
\be
\nabla_x^2\d(\vx,\tau)\,,
\ee
likewise derived from the Eulerian density field.

\subsection{Transfer function parameters}
\label{app:bao_param}

The parameters $r_{s, \mathrm{fid}}$, $A$, $\phi$, and $k_{\rm D}$ of the transfer function in \refeq{fdef} and the normalization of the smooth power-spectrum are determined by the best-fit to the linear power spectrum given by CLASS \cite{CLASS} in the fiducial cosmology (see \cite{Babic:2022dws} for details). For the field-level and standard post-reconstruction power-spectrum analysis of the SNG halo sample, we use $r_{s, \mathrm{fid}}=99.42\Mpch$, $A=0.07$, $\phi=0.67$, and $k_{\rm D}=0.19\iMpch$. For the Uchuu halo sample, we only fit $r_{s, \mathrm{fid}}$ and the normalization of the smooth power-spectrum, while we keep $A$, $\phi$, $k_{\rm D}$ fixed to the same values obtained for SNG. We obtain $r_{s, \mathrm{fid}}=96.50\Mpch$. The smooth power-spectrum normalization obtained in these fits cancels out when taking the transfer functions ratio in \refeq{fdef}, hence its value does not impact the final analysis and we do not quote it here. Note further that the linear power spectrum of the fiducial cosmology is always recovered exactly in the limit $r_s/r_{s,\rm fid}\to 1$, by construction of \refeq{fdef}.

\section{Standard BAO reconstruction analysis}
\label{app:reco}

Our reconstruction pipeline follows widely adopted approaches \cite{Rec_Eisenstein_2007}. To ensure a fair comparison with the field-level approach and guarantee that both methods have access to the same $k$-modes, we carefully select the smoothing scale and grid sizes employed in the reconstruction, as described in \cite{babic/etal:2024}. The choice of the smoothing scale is particularly important, given that the field-level approach employs a sharp-$k$ filter, while standard reconstruction involves a Gaussian filter. Here, we adopt $R = \Lambda^{-1}$ for the Gaussian smoothing scale. We consider this to be a conservative choice for the comparison, as it allows for a significant contribution from modes with $k > \Lambda$ in the standard reconstruction approach, while these modes are excluded from the field-level analysis.

\begin{figure}[t]
        \centering
            \includegraphics[width=0.5\linewidth]{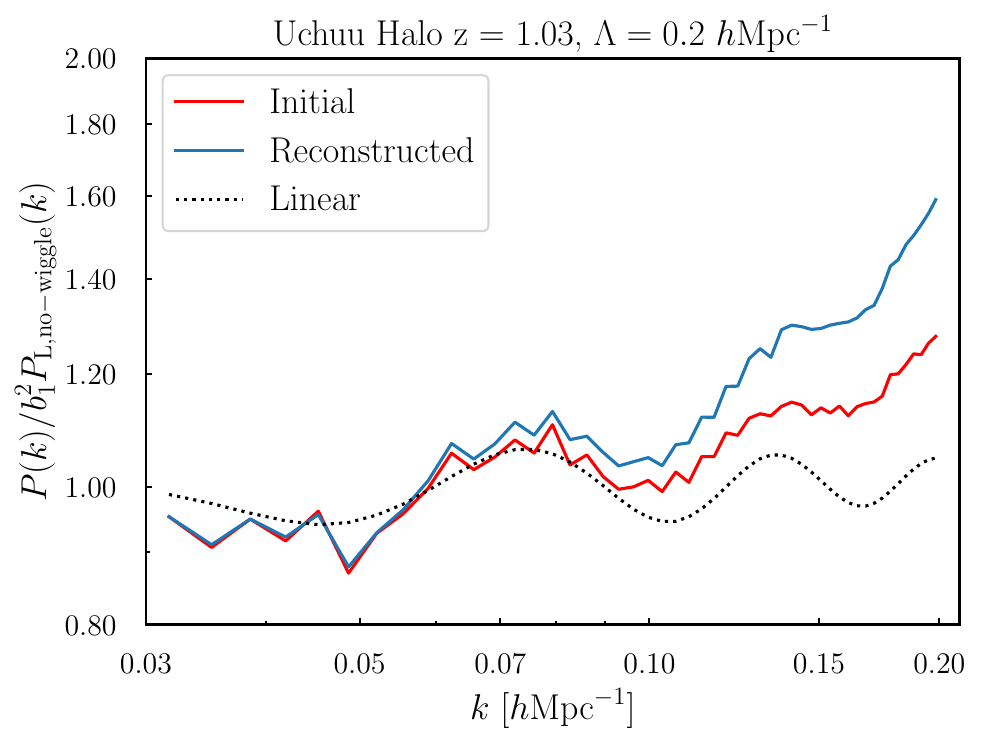}
            \caption{
            Same as \reffig{Pk-over-PLsm}, but for the Uchuu sample, showing similar trends as the SNG halo sample. The post-reconstruction broad-band enhancement is somewhat lower than in the SNG halo sample case, but the sample has higher signal-to-noise overall due to the higher number density. This higher signal-to-noise likely leads to the greater improvement in BAO scale precision seen for this sample for the highest cutoff, $\L=0.2\iMpch$.}
            \label{fig:Pk-over-PLsm-Uchuu}
 \end{figure}

Briefly, the standard reconstruction for a given cutoff $\L$ proceeds as follows:
\begin{enumerate}

    \item The tracers are assigned to a grid of size $ N_{\rm assign} = 3/2N^{\Lambda}$, where $N^\Lambda$ denotes the size of a grid that has support up to $k_{\rm Ny}=\Lambda$, using the NUFFT assignment scheme to obtain the pre-reconstruction tracer field $\delta_g$.

    \item The density field $\delta_g$ is smoothed using a Gaussian filter $W_R(k)$ with a smoothing scale $R = \L^{-1}$: $\delta_g(\bm{k}) \rightarrow W_R(k)\delta_g(\bm{k})$. 

    \item Using the smoothed density field, we find the estimated displacement $\boldsymbol{\psi}$, defined as 
      \begin{equation}
        \boldsymbol{\psi}(\bm{k}) \equiv -i\frac{\bm{k}}{k^2} \frac{W_R(k)\delta_g(\bm{k})}{b_\delta},
        \label{eq:psi}
      \end{equation}
where we employ a fixed value for $b_\delta$ estimated from the halo power spectrum. Specifically, $b_\delta^2$ is estimated as the ratio of the halo power spectrum on large scales and linear matter power spectrum at the same redshift.

    \item We interpolate $\boldsymbol{\psi}$ to find its value at the position of each tracer and use it to move the tracers.

    \item Once all the tracers have been shifted, we again use the NUFFT assignment scheme and a grid of size $3/2N^{\Lambda}$ to obtain the ``displaced'' density field $\delta_d$. Notice that this operation removes a large fraction of the large-scale perturbations in $\delta_g$.
 
    \item We generate a spatially uniform grid of particles and shift them by $\boldsymbol{\psi}$ to create the ``shifted'' field $\delta_s$, using the same grid size and assignment scheme as for $\delta_d$.
      
    \item The ``reconstructed density field'' is obtained as $\delta_g^{\text{rec}} = \delta_d - \delta_s$, where the field $-\delta_s$ re-instates the large-scale perturbations removed from $\delta_d$.
      
    \item Finally, we measure the power spectrum of the reconstructed field, $P_{\text{p-rec}}$, with $k_{\mathrm{max}} = \Lambda$.
\end{enumerate}
The result is illustrated in \reffig{Pk-over-PLsm} in the main text (for the SNG halo sample) and  \reffig{Pk-over-PLsm-Uchuu} below (for the Uchuu sample). Post-reconstruction, the BAO wiggles become noticeably more pronounced. 

Next, we determine the BAO scale parameter $r_s$, parametrized via the parameter $\beta \equiv r_s/r_{s,\rm fid}$, and its uncertainty by performing an MCMC analysis using the \texttt{emcee} sampler \cite{emcee_2013}. The analysis jointly infers the BAO scale with broad-band power spectrum fitting parameters, using  the following template:
  \begin{equation}
    P_{\rm model}(k|r_s) = (B_1 + B_2 k^2) P_{\rm L}(k|r_s) + A(k),
    \label{eq:Pk_model}
   \end{equation}
where $B_1$ and $B_2$ are fitting parameters which, together with $A(k)$, describe the broad-band shape of the post-reconstruction power spectrum, and
\begin{equation}
  P_{\rm L}(k|r_s) = f^2(k|r_s) P_{\rm L}^{\rm fid}(k)
\end{equation}
represents the linear power spectrum with the rescaled BAO feature, as introduced in \refeq{deltalin_shat}.
The term $A(k)$ is modeled as a third-order polynomial:
\begin{equation}
  A(k) = a_0 + a_2 k^2 + a_3 k^3.
    \label{eq:A_model}
\end{equation}
To compute the power spectrum, we apply linear binning and perform the fit in the range  $0.03\iMpch < k < \Lambda$.

The inference uses the following likelihood:
\begin{equation}
  -2\log\mathcal{P}_{P_{\text{p-rec}}} (r_s) = \sum_{k_i} \frac{\left(P_{\text{p-rec}}(k_i) - P_{\text{model}}(k_i|r_s)\right)^2}{\text{Cov}[P_{\text{best-fit}}(k_i)]} + {\rm const},
  \label{eq:L_P}
\end{equation}
where the sum is over the linear bins in wavenumber with central values $k_i$, and $P_{\text{p-rec}}(k)$ represents the post-reconstruction power spectrum of the  data.
It is worth noting that $P_{\text{model}}(k|r_s)$ already incorporates the stochastic contribution $P_\epsilon$ via the parameter $a_0$.
Following current standard analyses, we use the Gaussian covariance
\begin{equation}
    \text{Cov}[P_{\text{best-fit}}(k_i)] = 2\frac{ \left[P_{\text{best-fit}}(k_i,r_{s,\rm best-fit})\right]^2 }{m_i},
\end{equation}
where $m_i$ is the number of modes in the bin $i$. Before the MCMC analysis, the best-fit power spectrum, $P_{\text{best-fit}}$, is obtained by applying a curve-fitting procedure to determine the best-fit parameter values for the model function given the data $P_{\rm{data}}$. We stress that these parameters are only used to determine the covariance in the likelihood \refeq{L_P}, which is then kept fixed.
Since the covariance is constant (independent of all parameters that are varied), the normalization constant in \refeq{L_P} is irrelevant. 
For each $\Lambda$ value and halo sample, we run 32 chains with 6000 samples which give us around 2000 effective samples.
An example of the resulting posterior is shown in \reffig{Corner_plot_rec}.
We find that the model is a good fit to the post-reconstruction power spectrum; in case of the SNG sample at $\Lambda=0.2\iMpch$, for example, we find a reduced-$\chi^2$ of $1.01$ ($N_{\rm dof}=44$).

\section{Priors}
\label{app:Priors}

\subsection{Field-level inference priors}

In the FLI analysis, we assume the following priors:
\begin{gather*}
\P(r_s) = \U(0.9,\ 1.1)\times r_{s,\rm fid}, \\
\P(b_\delta) = \N(1.0,\ 5.0), \\
\P(b_O) = \N(0.0,\ 1.0),\quad O \notin \{ \d,\, \nabla^2\d \}, \\
\P(b_{\nabla^2\delta}) = \N(0.0 \Mpch,\ 5.0 \Mpch), \\
\P(\sigma_{\eps,0}) = \U(0.9\sigma_{\eps,\mathrm{Poisson}},\ 3.0\sigma_{\eps,\mathrm{Poisson}}), \quad \P(\sigma_{\eps,k^2}) = \U(-0.2 \Lambda^{-2},\ 5 \Lambda^{-2}),\numberthis\label{eq:FLI_priors}
\end{gather*}
where $\U(a,b)$ represents a uniform distribution between the lower bound $a$ and upper bound $b$, while $\N(\mu,\hat\sigma)$ represents a normal distribution with mean $\mu$ and standard deviation $\hat\sigma$. The third line lists the prior on all nonlinear bias parameters corresponding to the bias fields in \refeq{biaslist}. Only $b_\delta$ is sampled explicitly, while all other bias coefficients are analytically marginalized.

$\sigma_{\eps,\mathrm{Poisson}}$ in \cref{eq:FLI_priors} is the expected (Gaussian) RMS noise amplitude based on the Poisson shot noise expectation for the given cutoff, given the comoving number density of the tracers considered. As in the previous study \cite{fbisbi}, it is necessary to impose an informative lower limit on the noise level; without this limit, the FLI MCMC chains tend to get stuck in a region of unphysically low $\sigmaEps$. The cause of this drift is the incomplete noise treatment inherent in \refeq{FLI_likelihood}, which assumes perfectly Gaussian noise. Unfortunately, a consistent incorporation of non-Gaussian and density-dependent noise is still in a preliminary stage. However, we have verified that the lower limit of the prior on $\sigmaEps$ does not impact the posterior for the BAO scale noticeably.

The priors on the bias coefficients $\{ b_O \}$ are only weakly informative (and essentially uninformative in case of $b_\delta$). This has been verified in the context of the $\sigma_8$ inference
in \cite{Beyond2pt:preprint,fbisbi}. Since $r_s$ shows much weaker correlation with bias coefficients than $\sigma_8$, the bias priors are expected to have even less of an impact for the results presented here.

\subsection{Standard reconstruction analysis priors}

\begin{figure}[t]
        \centering
            \includegraphics[width=0.85\linewidth]{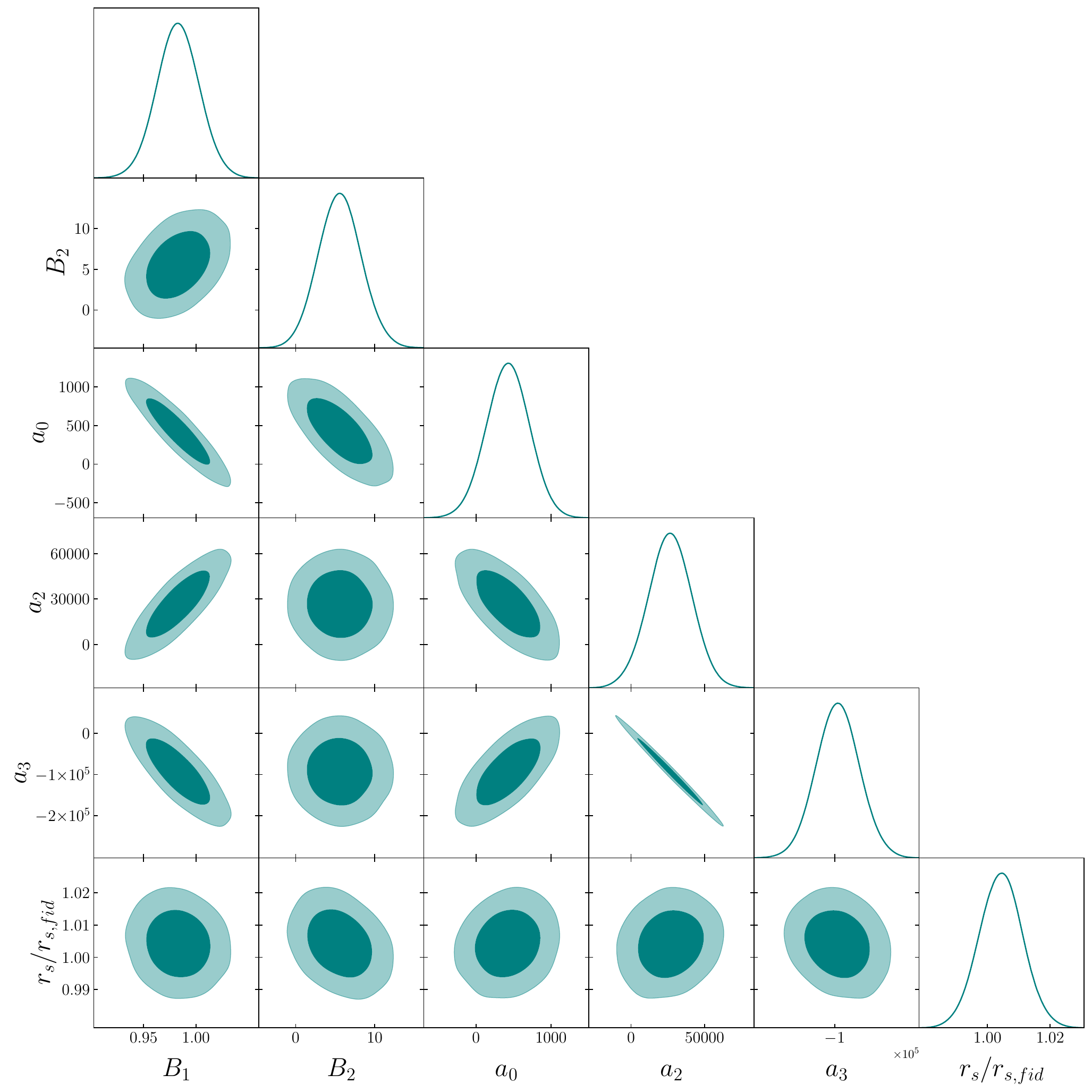}
            \caption{
            Parameter posterior of the post-reconstruction power spectrum analysis on the Uchuu sample for $\L=0.20 \iMpch$. The broad-band parameters $\{B_n, a_m\}$ are defined in \refeqs{Pk_model}{A_model}. The BAO scale parameter ($r_s / r_{s,\mathrm{fid}}$) shows very weak correlation with the broad-band parameters.}
            \label{fig:Corner_plot_rec}
 \end{figure}

For $r_s$, we adopt a uniform prior, $\mathcal{P}(r_s) = \mathcal{U}(0.6, 1.4)\times r_{s,\rm fid}$. For the remaining parameters in Eqs.~\eqref{eq:Pk_model}--\eqref{eq:A_model}, we adopt broad Gaussian priors.
The prior mean of the parameters $\boldsymbol{\theta}=\{B_1, B_2, a_0, a_2, a_3\}$ is set using the best-fit value obtained from a curve fit routine in Python, while the prior standard deviation $\hat\sigma_\theta$ is chosen individually for each parameter to ensure an uninformative prior range, where $\{\hat\sigma_{B_1},\hat\sigma_{B_2},\hat\sigma_{a_0},\hat\sigma_{a_2},\hat\sigma_{a_3}\} = \{10, 10^2, 10^4, 10^6, 10^6\}$.
We have verified that the inferred posterior on the BAO scale is insensitive to these prior choices, thanks to the fact that it is only weakly correlated with the broad-band parameters as can be seen from Fig.~\ref{fig:Corner_plot_rec}.

\section{Field-level inference posterior validation}
\label{app:validation}

An MCMC convergence analysis is particularly important in the case of high-dimensional inferences such as the FLI performed here. For this purpose, we employ two different strategies to initialize an MCMC chain: we initialize one MCMC chain from the true initial conditions (true-phase initialization, TPI) $\shat_{\mathrm{true}}$, and multiple chains at different random initial conditions and starting parameters (random-phase initialization, RPI) $\shat$. In case of RPI chains, we first perform a warm-up stage with initial conditions sampling at fixed parameter values.
Monitoring and comparing behaviors between TPI and RPI chains help us verify that the chains have converged. In case of the Uchuu sample, where ground-truth initial conditions are not available, we only have RPI chains starting from independent initial points in parameter space.

\begin{figure}[t]
   \centering
 \includegraphics[width=0.49\linewidth]{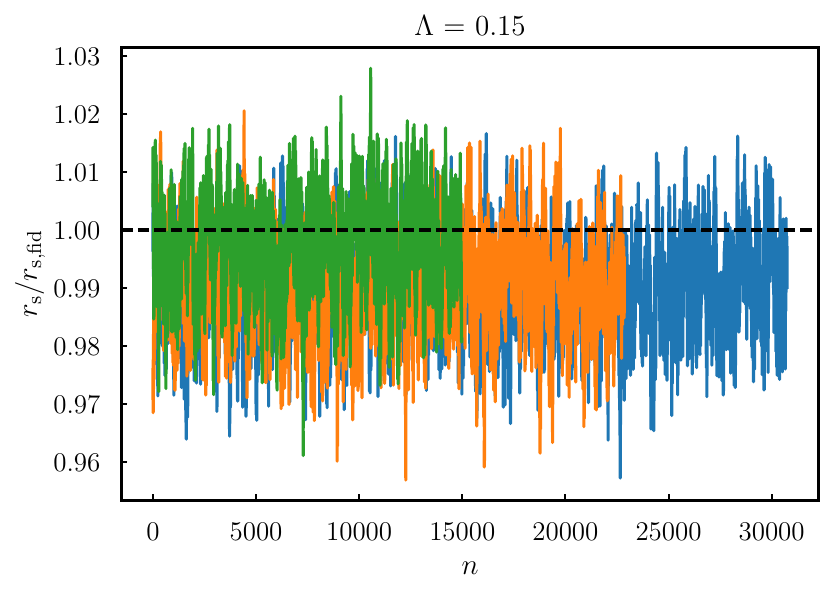}
 \includegraphics[width=0.49\linewidth]{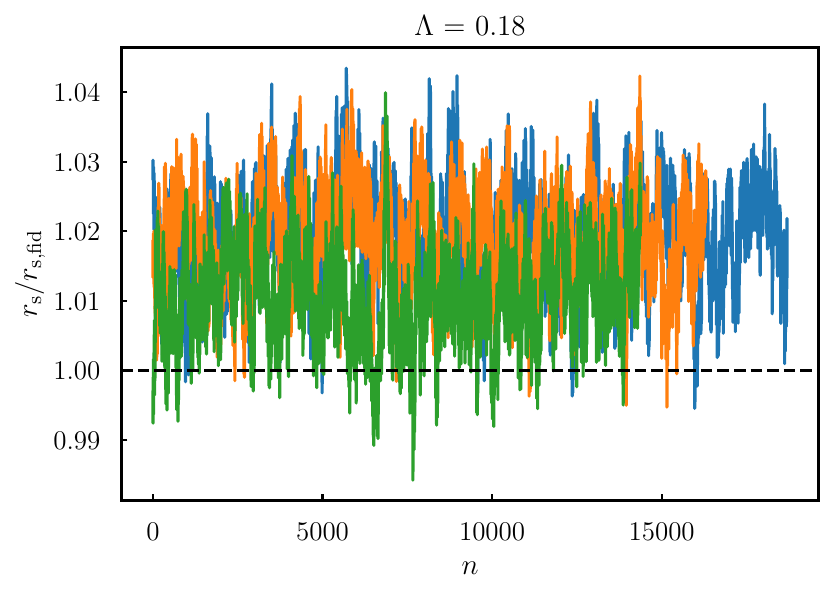}
 \includegraphics[width=0.49\linewidth]{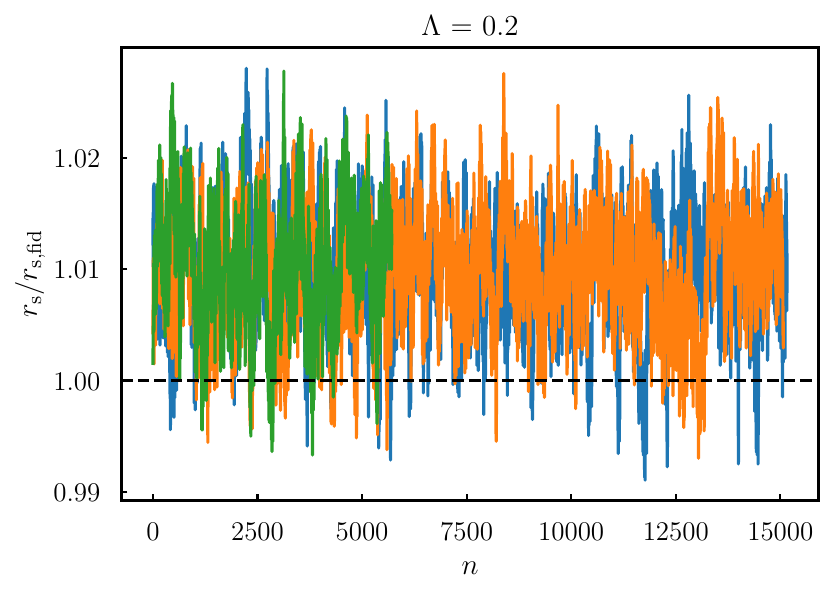}
   \caption{Parameter traces (as function of sample index $n$) of the scaled BAO sound horizon for TPI and RPI chains of the SNG tracer sample. The different panels show the different cutoffs considered. Different colors indicate chains starting at distinct initial points in parameter space.}
   \label{fig:beta_trace_SNG}
 \end{figure}

\begin{figure}[t]
   \centering
 \includegraphics[width=0.49\linewidth]{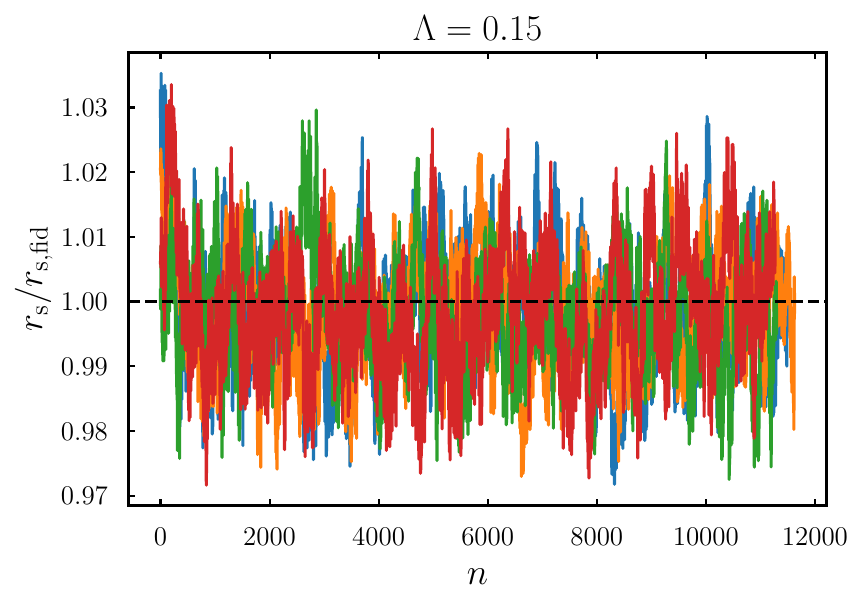}
 \includegraphics[width=0.49\linewidth]{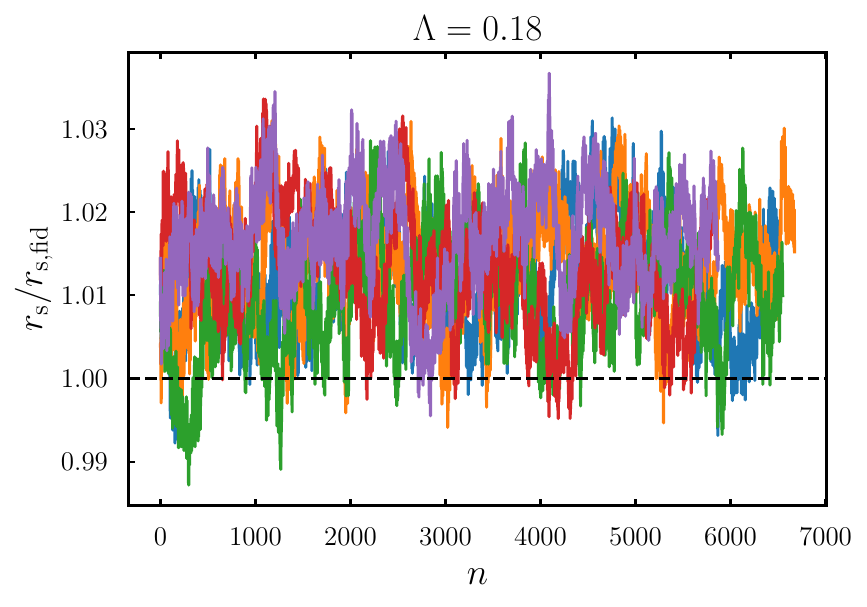}
 \includegraphics[width=0.49\linewidth]{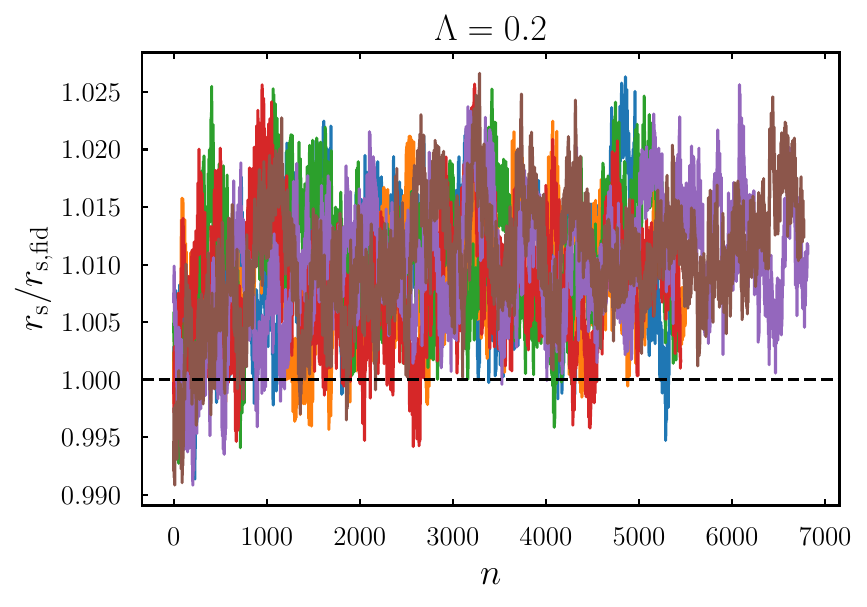}
   \caption{Parameter traces (as function of sample index $n$) of the scaled BAO sound horizon for the RPI chains of the Uchuu tracer sample. The different panels show the different cutoffs considered. Different colors indicate chains starting at distinct initial points in parameter space.}
   \label{fig:beta_trace_Uchuu}
 \end{figure}

\begin{figure}[t]
   \centering
 \includegraphics[width=0.4\linewidth]{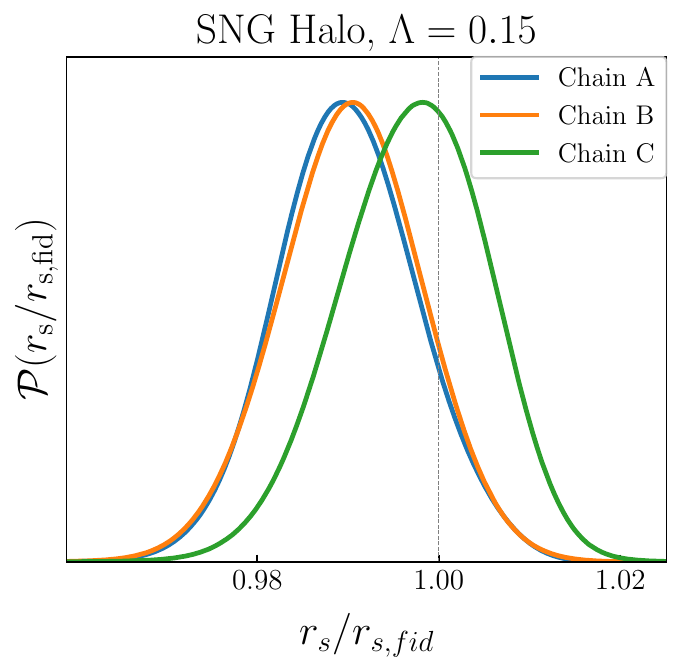}
 \includegraphics[width=0.4\linewidth]{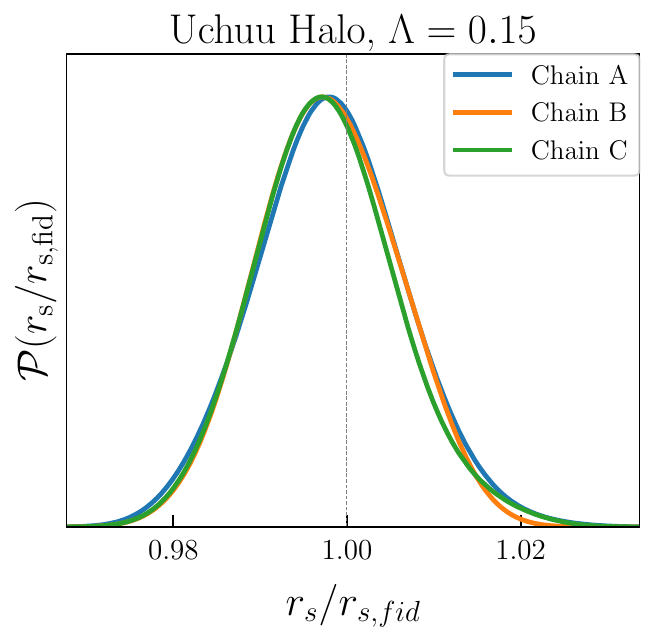}
   \caption{Posterior distributions of $r_{\rm s}/r_{\rm s, fid}$ from three FLI chains with different initial conditions are shown for both SNG halos (left) and Uchuu halos (right). In both cases, the chains converge to consistent results, indicating the robustness of the inference. Note that Chain C in the left panel is considerably shorter than Chains A, B (cf. \reffig{beta_trace_SNG}).}
   \label{fig:beta_posterior_SNG}
 \end{figure}

\reffig{beta_trace_SNG} shows parameter traces from one TPI and two RPI chains each for the different cutoffs considered on the SNG tracer sample (the warm-up stage of the RPI chains is not shown here). \reffig{beta_trace_Uchuu} shows the same for the Uchuu sample.
After a brief ``burn-in'' phase, the chains show very similar statistical behavior, all sampling the underlying posterior as can be seen from the left panel in \reffig{beta_posterior_SNG}.

\begin{figure}[t]
        \centering
            \includegraphics[width=0.85\linewidth]{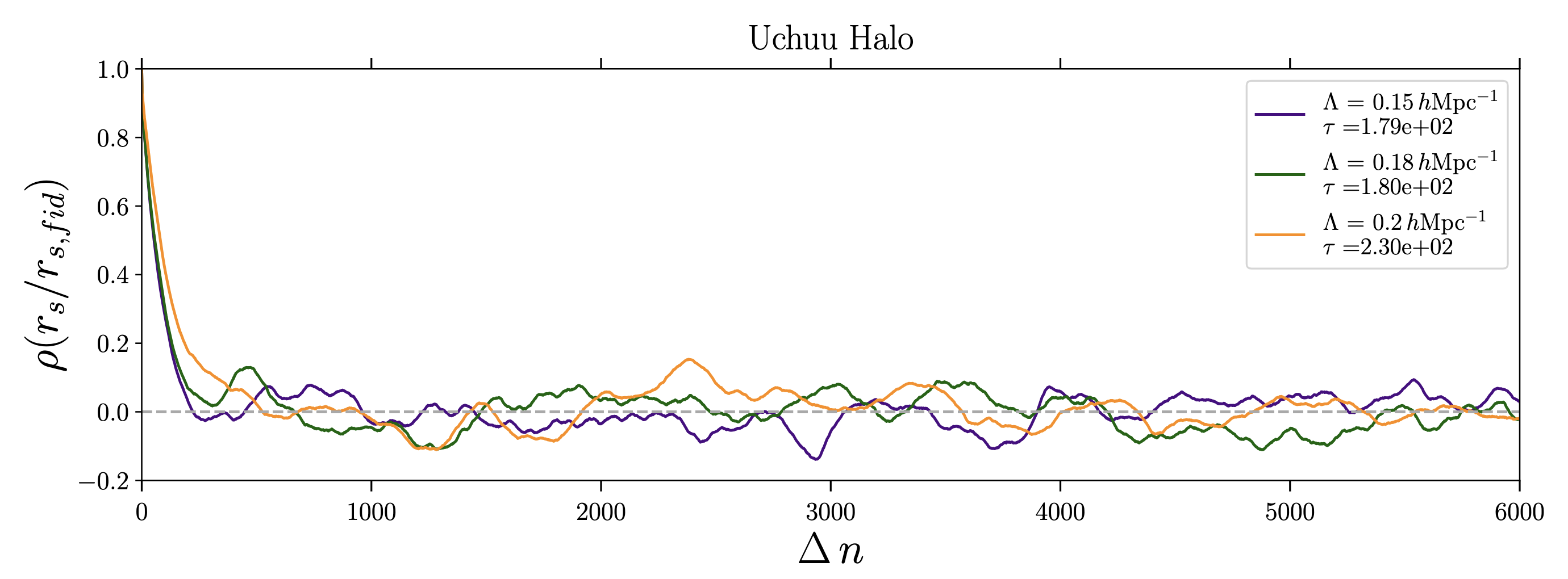}
            \caption{Normalized auto-correlation function of the BAO scale parameter inferred from the combined MCMC chains run on Uchuu halos for each value of $\Lambda$. For better visibility, we restrict to a sample lag of $\Delta n \leqslant 6000$. The estimated correlation length $\tau$ is indicated, demonstrating that the chains exhibit autocorrelation over roughly 200 samples for all $\Lambda$ values.}
            \label{fig:uchuu_autoco}
 \end{figure}

Given our focus on possible improvements in the precision of the BAO inference, we pay particular attention to validating the posterior width for $r_s$. 
The MCMC sampling error on this quantity can be estimated as $\sigma(r_s)/\sqrt{\mathrm{ESS}}$ where ESS is the effective number of independent samples, or effective sample size of $r_s$. \cref{tab:EFT-fieldlevel_neff} reports the ESS for each FLI analysis shown in the text. Additionally, Fig. \ref{fig:uchuu_autoco} shows the normalized autocorrelation function of the BAO scale parameter for the Uchuu halos, from which we observe that the correlation length $\tau$ is only a few hundred samples.

\begin{table}
    \centering
    \begin{tabular}{l|c|cccc}
        \hline
        \hline
        Sample & Cutoff $\L=\kmax$ & $r_s^{\rm mean}/r_{s,\rm fid}$ & $r_s^{\rm MAP}/r_{s,\rm fid}$ & $\sigma(r_s)/r_{s,\rm fid}$ & ESS [samples] \\
        \hline
    SNG & $0.15$ & 0.9902 & 0.9920 & 0.0077 & 735 \\
    SNG & $0.18$ & 1.0192 & 1.0197 & 0.0073 & 145 \\
    SNG & $0.20$ & 1.0103 & 1.0097 & 0.0052 & 237 \\
        \hline   
    Uchuu & $0.15$ & 0.9983 & 0.9975 & 0.0085 & 421\\
    Uchuu & $0.18$ & 1.0117 & 1.0126 & 0.0065 & 241 \\
    Uchuu & $0.20$ & 1.0103 & 1.0100 & 0.0048 & 230 \\
        \hline   
        \hline
    \end{tabular}
    \caption{
    Summary of results of the FLI inference on the two halo samples, each for three different cutoffs. Starting from the third column, we give the mean and MAP values for $r_s$, the 68\% CL error bar relative to $r_{s,\rm fid}$, and the effective sample size ESS, all for the combined MCMC chains.
    }
    \label{tab:EFT-fieldlevel_neff}
\end{table}

\clearpage
\section{Where does the BAO information come from?}
\label{app:Fisher}

In order to better understand the source of BAO information in the field-level
and standard reconstruction-based BAO inference procedures, we consider
a simplified setup which allows us to make analytical progress.
We ignore the stochasticity (shot noise) in the data, and so assume that
the data is infinitely informative up to the maximum scale $\kmax=\Lambda$
included in the analysis. 
As shown in \cite{Cabass:2023nyo,MAP_paper}, the posterior
for FLI can then be written as
\ba
-2\ln \P_{\rm FLI}[\bOset, r_s | \d_g] =\:& 
\sum_{\vk}^\L\frac{|\dgdet^{-1}[\d_g,\bOset](\vk)|^2}{\Plin(k|r_s)} 
+ 2\ln \left| \frac{\Del \dgdet}{\Del \dlinL}  \right|_{\dgdet^{-1}[\d_g,\bOset]} \vs
& + \ln \left[\prod_{\vk}^\L 2\pi \Plin(k|r_s)\right] - 2\ln \P_{\rm prior}(\bOset, r_s) + {\rm const}\,
,
\label{eq:logpost}
\ea
where $\bOset$ denotes the set of bias parameters, and we have specialized the set of cosmological parameters to $r_s$. Further, $\dgdet^{-1}$ denotes the formal inverse of the forward model. In the following, we assume uninformative priors on $r_s$, while we keep $\{b_O\}$ fixed, so that we can drop $\P_{\rm prior}$.

Now we take the derivative of \refeq{logpost} with respect to $r_s$:
\ba
\frac{\partial}{\partial r_s} \left(-\ln \P_{\rm FLI}[\bOset, r_s | \d_g]\right) =\:& 
\frac12 \sum_{\vk}^\L\left[\Plin(k|r_s) - |\dgdet^{-1}[\d_g,\bOset](\vk)|^2
  \right] \frac1{[\Plin(k|r_s)]^2}\frac{\partial \Plin(k|r_s)}{\partial r_s}\,.
\label{eq:dlogpostdlambda}
\ea
Setting this relation to zero, we see that, at fixed values of the bias parameters $\bOset$, the maximum-a-posteriori (MAP) point for the power spectrum parameters corresponds to the point
where the mismatch between the maximum-likelihood estimator of the linear density field, $\dgdet^{-1}[\d_g,\bOset]$, and the expection $\Plin(k|r_s)$ vanishes, or more precisely, has no overlap with the gradient of $\Plin$ with respect to $r_s$.
Note that $\partial\Plin/\partial r_s$ is an oscillatory function (i.e. it has no broad-band part).
In the following, we will assume that the $b_O$ are set to their MAP values,
which in the absence of model mismatch are the ground-truth values. In
the actual FLI analysis, we of course vary both $\bOset$ and $r_s$
at the same time.

We now derive the Fisher information on the BAO scale parameter $r_s$.
Using that
\be
\< \Plin(k|r_s) - |\dgdet^{-1}[\d_g,\bOset](\vk)|^2 \> = 0,
\ee
we obtain
\ba
F_{r_sr_s}^{\rm FLI} = -\<\frac{\partial^2}{\partial r_s^2} \ln \P_{\rm FLI}[\bOset, r_s | \d_g]\> 
   =\:& 
\frac12 \sum_{\vk}^\L\frac1{[\Plin(k|r_{s,\rm fid})]^2} \left(\frac{\partial \Plin(k|r_{s,\rm fid})}{\partial r_{s,\rm fid}}\right)^2\,.
\label{eq:FisherFLIapp}
\ea
This is in fact precisely the Fisher information on $r_s$ contained in the \emph{linear power spectrum up to the scale $\L$}. Thus, the information in the field-level posterior on
the linear power spectrum parameters matches exactly that in the linear
density field, if the bias parameters are perfectly known. This is just another
statement that the Bayesian field-level analysis is optimal in this case.

The post-reconstruction power spectrum likelihood, on the other hand, is \refeq{L_P}, or
\ba
-2\ln \P_{P_{\text{p-rec}}}[r_s | \d_g] =\:&
\sum_{\vk}^\L
\bigg[|\dgrec[\d_g](\vk)|^2 - P_{\text{model}}(k|r_s)\bigg]^2
\frac{1}{{\rm Var}[ P_{\text{model}}(k|r_{s,\rm fid})]}
+  {\rm const}\,,
\label{eq:L_P2}
\ea
where we have not written the $r_s$-independent normalization.
Here, we have summed over individual modes of $|\dgrec(\vk)|^2$ to emphasize the similarity with the FLI case, while in practice and in \refeq{L_P} one sums over finite and angle-averaged bins in $|\vk|$.
With an analogous derivation to that leading to \refeq{FisherFLI},
and assuming that $P_\text{model}(k) = \langle P_{\text{p-rec}}(k) \rangle$, we obtain
\ba
F_{r_sr_s}^{\rm rec-P(k)} = 
-\<\frac{\partial^2}{\partial r_s^2} \ln \P_{\rm rec-P(k)}[r_s | \d_g]\> 
   =\:&
\sum_{\vk}^\L
\frac{1}{{\rm Var}[ P_{\text{p-rec}}(k|r_{s,\rm fid})]}
\left(\frac{\partial P_{\text{p-rec}}(k|r_s)}{\partial r_s}\right)^2 \vs
   =\:&
\frac12 \sum_{\vk}^\L
\frac{1}{[ P_{\text{p-rec}}(k|r_{s,\rm fid})]^2}
\left(\frac{\partial P_{\text{p-rec}}(k|r_s)}{\partial r_s}\right)^2\,.
\label{eq:FisherPkrecapp}
\ea
Here, we have assumed a Gaussian covariance for the post-reconstruction power spectrum, so that for a single mode, ${\rm Var}[ P_{\text{p-rec}}(k)] = 2 [ P_{\text{p-rec}}(k)]^2$.

\end{document}